\documentstyle[11pt,a4]{article}
\newcommand{\mbf}[1]{\mbox{\boldmath$#1$}}

\begin{document}
\title{Feynman versus Bakamjian-Thomas in Light Front Dynamics}
\author{ W.R.B. de Ara\'ujo$^a$, M. Beyer$^b$,
T. Frederico$^c$ and
  H.J. Weber$^b$\footnote{permanent address: Dept. of    Physics,
    University of     Virginia, Charlottesville, U.S.A.}\\\small
$^a$Laborat\'orio do Acelerador Linear, Instituto de F\'\i sica
da USP\\\small C.P. 663118, CEP 05315-970, S\~ao Paulo, Brazil\\\small
$^b$ FB Physik, Universit\"at Rostock, 18051 Rostock, Germany\\\small
$^c$ Dep. de F\'\i sica, Instituto Tecnol\'ogico de Aeron\'autica,
Centro T\'ecnico Aeroespacial, \\\small
12.228-900 S\~ao Jos\'e dos Campos, S\~ao Paulo, Brazil.}
\date{}
\maketitle
\begin{abstract}
  We compare the Bakamjian-Thomas (BT) formulation of relativistic
  few-body systems with light front field theories that maintain
  closer contact with Feynman diagrams. We find that Feynman diagrams
  distinguish Melosh rotations and other kinematical quantities
  belonging to various composite subsystem frames that correspond to
  different loop integrals. The BT formalism knows only the rest frame
  of the {\em whole} composite system, where everything is evaluated.
\end{abstract}

\vspace{3ex}
The goal of this article is to point out subtle, but
important, differences between the Bakamjian-Thomas (BT) formulation
of relativistic few-body systems~\cite{BT53} and light front field
theories that maintain closer contact with Feynman diagrams (see,
e.g., Ref.~\cite{karmanov}). To be specific, we consider the triangle
diagram that is a major ingredient of recent electromagnetic and weak
baryon form factor evaluations in light front dynamics.

We start with the effective Lagrangian for the N-q coupling
\begin{eqnarray}
{\cal{L}}_{N-3q}=\sum_{\left\{ i,j,k\right\} }\overline{\Psi}_{(i)}
i\tau _2\gamma _5\Psi_{(j)}^C\overline{\Psi} _{(k)}\Psi _N
\label{lag}
\end{eqnarray}
where $\tau _2$ is the isospin matrix and the sum is over permutations
of $\{1,2,3\}$. The conjugate quark field is $\Psi^C=C
\overline{\Psi}^\top $, where $C=i\gamma^2\gamma^0$ is the charge
conjugation matrix.

In the nucleon rest frame, and choosing $i=1$, $j=2$, and $k=3$ the
spin coupling of the quarks to the nucleon is given by:
\begin{equation}
\chi (s_1,s_2,s_3;s_N)=\overline{u}_{1}
\gamma _5 u_{2}^C\;
\overline{u}_{3}u_{N} \ ,
\label{nuc}
\end{equation}
where the light-front spinor $u_1=u(p_1,s_1)$ is
\begin{eqnarray}
u(p,s)=\frac{p\cdot\gamma+m}{\sqrt{2 p^+ 2 m }}
\gamma^+ \gamma^0
\left(\begin{array}{c}
\chi_{s} \\0\end{array}
\right)  \
\label{lf}
\end{eqnarray}
where $p^+=p^0+p^3$, $p^-=p^0-p^3$, $\mbf{p}_\perp=(p_1,p_2)$, and
$\chi_{s}$ is the two component Pauli spinor. The Dirac spinor of the
instant form
\begin{eqnarray}
u_{D}(p,s)=\frac{p\cdot\gamma+m}{\sqrt{2 m(p^0+m) }}
\left(\begin{array}{c}
\chi_{s} \\
0
\end{array}
\right)  \
\label{dirac}
\end{eqnarray}
carries the subscript $D$. The expression Eq.~(\ref{nuc}) appears in
the evaluation of the two-loop Feynman diagram of the $J^+=J^0+J^3$
component of the nucleon electromagnetic current once the integrations
over the $'-'$ components of the quark momenta, $p_i^- = p_i^0 -
p_i^3$, are carried out. There are two loops rather than one because
of the integrals over both relativistic Jacobi relative momentum
variables $q_3, Q_3$ defined as usual ($x_i=p^+_i/P^+$)
\begin{eqnarray}
p_1&=&q_3-{x_1\over 1-x_3}Q_3+x_1 P,\label{reco1}\\
p_2&=&-q_3-{x_2\over 1-x_3}Q_3+x_2 P,\label{reco2}\\
p_3&=&Q_3+x_3 P,\
\label{reco3}
\end{eqnarray}
and $P=p_1+p_2+p_3$ all valid for the $+$ and $\perp$ components only,
 so that $q_3^+=0,\ Q_3^+=0$.
Equivalently, there are integrations over three quark momentum
variables in the nucleon rest frame, $k_i=p_i-x_i P,$ with the
restrictions $\sum_i \mbf{k}_{\perp i}=0$ and $\sum_i x_i=1.$

This spin-flavor invariant of the nucleon with quark pair spin zero is
the simplest of a basis of 8 such states given in greater detail in
Ref.~\cite{BKW98}, for example. The only nucleon spin invariant used
and tested in form factor calculations contains the additional
projector $\gamma \cdot P+M_0$ onto large Dirac components, a
characteristic feature of the BT formalism, where $P$ is the total
nucleon momentum and $M_0^2$ the sum of the free quark light cone
energies.

The residues of the triangle Feynman diagram are evaluated at the
on-$k^-$-shell poles of the spectator particles~\cite{TF}. The
numerator of the fermion propagator of the quark which absorbs the
photon momentum can be considered on-$k^-$-shell because
$(\gamma^+)^2=0$. More generally, spin sums may be performed
covariantly provided they occur before the $k^-_i$ integrations. Thus,
all the numerators of the fermion propagators can be substituted by
the positive energy spinor projector, written in terms of light-front
spinors.

The Melosh rotation is given by:
\begin{eqnarray}
\left[R_{M}(p)\right]_{s's}=\overline{u}_D(p,s')u(p,s)
\ .
\label{mel}
\end{eqnarray}

To evaluate Eq.~(\ref{nuc}), we observe that the Wigner rotation of
the light-front spinors is one for kinematical light-front boosts. Let
us recall that, as a result of the transitivity of the kinematic
generators in the front form, a wave function is defined everywhere,
once it is defined in the rest frame of the composite system. Thus,
the matrix element of the pair coupled to spin zero is evaluated in
the rest frame of the pair (cm) which, again, is found by a
kinematical light-front boost $\Lambda$ from the nucleon rest frame.
Because the Wigner rotation is unity for such a Lorentz
transformation, we can write (viz. ${u}_{\rm cm}(\vec k^{{\rm cm}},s)
= {u}(\vec k^{{\rm cm}},s)$):
\begin{eqnarray}
I(s_1,s_2,0)&=&\overline{u}(\vec k_1,s_1)
\gamma _5u^C(\vec k_2,s_2)\nonumber\\
&=& \overline{u}(\vec k^{{\rm cm}}_1,s_1)
\gamma_5 u^C(\vec k^{{\rm cm}}_2,s_2)
\ , \label{eqn:i}
\end{eqnarray}
where $\vec k^{{\rm cm}}=(k^{+{\rm cm}},\mbf{k}_\perp^{{\rm cm}})$ are
the kinematical momentum variables of each particle 1 or 2 in the rest
frame of the pair 12, $k^{{\rm
  (cm)}\mu}=(\Lambda k)^\mu$. The particle momenta in the pair rest
frame are
obtained by a kinematical light-front transformation from those in the
nucleon rest frame to the pair rest frame due to the transitivity of
kinematic generators mentioned above. Thus inserting the completeness
relation for positive energy Dirac spinors in Eq.~(\ref{eqn:i}), we
obtain:
\begin{eqnarray}
I(s_1,s_2,0)&=&\sum_{\bar s_1 \bar s_2}
\overline{u}(\vec k^{{\rm cm}}_1,s_1)u_D(\vec k^{{\rm cm}}_1,\bar s_1)
\overline u_D(\vec k^{{\rm cm}}_1,\bar s_1)\nonumber\\
&&\gamma _5\, C\, \overline{u}^\top_D(\vec k^{{\rm cm}}_2,\bar s_2)
\left(\overline{u}(\vec k^{{\rm cm}}_2,s_2)
u_D(\vec k^{{\rm cm}}_2,\bar s_2)\right)^\top
\ . \label{eqn:i2}
\end{eqnarray}
Using the definition of the Dirac spinors we get the relevant
Clebsch-Gordan coefficients,
\begin{eqnarray}
\overline u_D(\vec k^{{\rm cm}}_1,\bar s_1)\gamma _5 C
\overline{u}^\top_D(\vec k^{{\rm cm}}_2,\bar s_2)
\rightarrow\chi_{\bar s_1}^\dagger i\sigma_2\chi^*_{\bar s_2}
=\sqrt{2}\;\langle \frac{1}{2} \bar s_1  \frac{1}{2} \bar s_2|
00\rangle \ . \label{i3}
\end{eqnarray}

{}From Eqs.(\ref{nuc}), (\ref{mel}), (\ref{eqn:i2}) and (\ref{i3}), we
finally write the expression for the spin coupling of the nucleon and
the quarks, resulting from one part of the effective Lagrangian:
\begin{equation}
\chi (s_1,s_2,s_3;s_N)= \sum_{\bar s_1 \bar s_2}
\left[ R^\dagger_M(\vec k_1^{{\rm cm}})\right]_{s_1\bar s_1}
\left[ R^\dagger_M(\vec k_2^{{\rm cm}})\right]_{s_2\bar s_2}
\left[ R^\dagger_M(\vec k_3)\right]_{s_3 s_N}
\chi_{\bar s_1}^\dagger i\sigma_2\chi^*_{\bar s_2} \;
. \label{numel}
\end{equation}
The above expression of the nucleon spin wave-function differs from
the Bakamjian-Thomas construction in so far as rest frames of
composite subsystems play a role in Feynman diagrams, while in the BT
only the overall cms matters. In particular, the Melosh rotations of
the spin-zero coupled pair (12) have the momentum arguments evaluated
in the rest-frame of the pair in Eq.~(\ref{numel}), while in the BT
construction the arguments of the Melosh rotations are all evaluated
in the nucleon rest frame. Also, various total momentum $'+'$
components, such as $P^+_{12}$ and $P^+$ now appear in different
frames, whereas in the BT case only $M_0$ occurs for $P^+$ in the
nucleon rest frame.

To illustrate the different kinematics in the two-body c.m. system
(Feynman) and three-body frames (BT formulation) we compare the
energy of quark $1$, i.  e. $p_1\cdot (p_1+p_2)/M_2$ and
$p_1\cdot P/M_0$, where $M_2^2=(p_1+p_2)^2$ is the  mass
squared of the two-body $(12)$-subsystem and $M^2_0=P^2$ that of the
nucleon. Using $q_3^2=(x_2p_1-x_1p_2)^2/(1-x_3)^2$ to obtain $2p_1\cdot p_2$, 
we find 
\begin{eqnarray}
M_2^2 &=& \frac{1-x_3}{x_1x_2} (x_2m_1^2 + x_1m_2^2)
-\frac{(1-x_3)^2}{x_1x_2} q_3^2, \\
M_0^2&=&\sum_{i=1}^3 \frac{m_i^2}{x_i} -\frac{1-x_3}{x_1x_2} q_3^2
-\frac{Q_3^2}{x_3(1-x_3)}.
\end{eqnarray}
We are careful to define the relevant projections with
four-vectors whose '$+$' components are zero, viz.  $\pi_{12}\equiv
p_1+p_2-(1-x_3)P$ and $\pi_1\equiv p_1-x_1 P$, to avoid using off-shell 
'$-$' components of the momenta. Using
\begin{eqnarray}
\pi_1^2&=&\left(q_3-\frac{x_1}{1-x_3}\, Q_3\right)^2=
m_1^2+x_1^2M_0^2-2x_1p_1\cdot
 P,\\
\pi_{12}^2 &=&Q_3^2
=M^2_2-2(1-x_3)(p_1+p_2)\cdot P+(1-x_3)^2 M_0^2
\label{eqn:pi12}
\end{eqnarray}
and
\begin{eqnarray}
\pi_1\cdot\pi_{12}&=&p_1\cdot(p_1+p_2)-(1-x_3)p_1\cdot P
    \nonumber\\&&  -x_1P\cdot(p_1+p_2)+x_1(1-x_3)M^2_0\nonumber\\
&=& -q_3\cdot Q_3 +\frac{x_1}{1-x_3}\; Q_3^2
\end{eqnarray}
to eliminate $(p_1+p_2)\cdot P$ in eq.~(\ref{eqn:pi12}),
we arrive at
\begin{eqnarray}
p_1\cdot P       &=&      \frac{\left(m_1^2-q_3^2\right) }{2x_1}
+ \frac{ q_3\cdot Q_3}{1-x_3}
                + \frac{x_1}{2} \left(M_0^2-\frac{
Q_3^2}{(1-x_3)^2}\right),\\
p_1\cdot (p_1+p_2) &=& \frac{1-x_3}{2x_1}
\left(m_1^2-q_3^2\right) + \frac{x_1 M_2^2}{2(1-x_3)}.
\end{eqnarray}
Clearly, the momentum variables of the (12)-subsystem depend only on
$M_2$ and $q_3$, while those in the nucleon c.m. system also depend on
$M_0$ and $Q_3$. As a consequence we expect also dynamical quantities
to change, e.g. form factors.

The same considerations will apply to the pair-spin 0 invariant with
an additional $\gamma \cdot P$ from the projector which reduces to
$\gamma_0$ in the nucleon rest frame. Another instructive spin-flavor
invariant will be discussed next, where the boost $\Lambda$ appears
explicitly, because of the vector character.
\par
Let us now consider the vector spin-flavor coupling
\begin{eqnarray}
\chi (s_1,s_2,s_3;s_N)=\overline{u}_{1}
\gamma ^{\mu} u_{2}^C\;
\overline{u}_{3}\gamma _{\mu}\gamma _5 u_{N} \ ,
\end{eqnarray}
where the spins of the 12-pair are coupled to unity and the relevant
vector-isospin matrix element has been omitted for simplicity. Instead
of Eq.\ref{eqn:i2}, we now obtain the coupling
\begin{eqnarray}
I^\mu(s_1,s_2,1)&=&\overline{u}(\vec k_1,s_1)
\gamma ^{\mu } u^C(\vec k_2,s_2)\nonumber\\
 &=& \overline{u}(\vec k_1^{\rm cm},s_1)
(\Lambda^{-1} \gamma) ^{\mu } u^C(\vec k_2^{\rm cm},s_2)
\label{v}
\end{eqnarray}
Eq.~(\ref{v}) then leads to the
Clebsch-Gordan coefficients upon restricting to the large Dirac
components ($\chi_s = [1,0] u(k,s)$), viz.
\begin{eqnarray}
\lefteqn{
\overline u_D(\vec k^{{\rm cm}}_1,\bar s_1)\gamma ^{\nu } C
\overline{u}^\top_D(\vec k^{{\rm cm}}_2,\bar s_2)\;
\overline u_D(\vec k_3,\bar s_1)(\Lambda\gamma) _{\nu} \gamma _5 u_N}
\nonumber\\
& \rightarrow&
\chi_{\bar s_1}^\dagger
( \vec{\sigma} i\sigma_2)
\chi^*_{\bar s_2}\cdot
\chi_{\bar s_3}^\dagger \vec{(\Lambda\sigma)} \chi_{s_N}
\nonumber\\
&=&-\sqrt{6}\; \sum_{mm'} \Lambda_{m'm}
\langle \frac{1}{2} \bar s_1  \frac{1}{2} \bar s_2
|1m'\rangle\;
\langle \frac{1}{2} \bar s_3  \frac{1}{2}  s_N|1,-m\rangle ,\
\label{vi}
\end{eqnarray}
and to
\begin{eqnarray}
\chi (s_1,s_2,s_3;s_N)&=& \sum_{\bar s_1 \bar s_2 \bar s_3}
\left[ R^\dagger_M(\vec k_1^{{\rm cm}})\right]_{s_1\bar s_1}
\left[ R^\dagger_M(\vec k_2^{{\rm cm}})\right]_{s_2\bar s_2}
\left[ R^\dagger_M(\vec k_3)\right]_{s_3 \bar s_3}\nonumber\\
&&\qquad\qquad
\times\;\chi_{\bar s_1}^\dagger
\left(\vec \sigma i\sigma_2\right) \chi^*_{\bar s_2}
\,\cdot\,\chi_{\bar s_3}^\dagger \vec {(\Lambda\sigma)} \chi_{ s_N}
. \label{numelvec}
\end{eqnarray}
Other spin-flavor 3-quark couplings are treated similarly.

In conclusion, we compare the evaluation of Feynman diagrams to the BT
formulation of multi-quark systems. We emphasize that Feynman diagrams
distinguish Melosh rotations and other kinematical quantities
belonging to various composite subsystem frames that correspond to
different loop integrals.  Moreover, the light-cone spinors in Eqs.
(\ref{eqn:i}), (\ref{eqn:i2}), and (\ref{v}) are no longer all in the
nucleon rest frame, which has consequences for the {\em normalization}
of the spin-flavor invariants. This may become important at higher
momentum tranfers and is relevant for the orthogonality of the wave
functions (i.e. at $q^2=0$). The BT formalism knows only the rest
frame of the {\em whole} composite system, where everything is
evaluated.

Thus, BT is much closer to nonrelativistic few-body theory, apart from
ignoring systematically small Dirac components, so that one is
justified calling it 'minimally relativistic'.

{\bf Acknowledgements:} HJW and TF are grateful to G. R\"opke and the
many
particle research group for their warm hospitality. WRBA thanks
CNPq for  financial support and TF thanks
CNPq and CAPES/DAAD/PROBRAL.

\end{document}